\documentclass[twocolumn,showpacs,amsmath,amssymb,10pt]{revtex4}
\usepackage{amsfonts}
\usepackage{bm}

\newcommand{\ot}{{\,\otimes\,}}
\newcommand{{\Cd}}{{\mathbb{C}^d}}

\def\oper{{\mathchoice{\rm 1\mskip-4mu l}{\rm 1\mskip-4mu l}%
{\rm 1\mskip-4.5mu l}{\rm 1\mskip-5mu l}}}
\def\<{\langle}
\def\>{\rangle}

\begin{document}
\title{\textbf{New class of states with positive partial transposition}} \author{Dariusz
Chru\'sci\'nski and Andrzej Kossakowski\thanks{email:
darch@phys.uni.torun.pl} }
\affiliation{Institute of Physics, Nicolaus Copernicus University,\\
Grudzi\c{a}dzka 5/7, 87--100 Toru\'n, Poland}

\begin{abstract}

We construct a new class quantum states bipartite $d \ot d$ states
which are positive under partial transposition (PPT states). This
class is invariant under the maximal commutative subgroup of
$U(d)$ and contains as special cases many well known examples of
PPT states. States from our class provide new criteria for testing
indecomposability of positive maps. Such maps are crucial in
constructing entanglement witnesses.

\end{abstract}

\maketitle

 The interest on  quantum entanglement has
dramatically increased during the last two decades due to the
emerging field of quantum information theory \cite{QIT}. It turns
out that quantum entangled states may be used as basic resources
in quantum information processing and communication, such as
quantum cryptography, quantum teleportation, dense coding, error
correction, and quantum computation.

A fundamental problem in quantum information theory is to test
whether a given state of a composite quantum system is entangled
or separable. Several operational criteria have been proposed to
identify entangled states \cite{criteria}.  The most famous
Peres-Horodecki criterion \cite{Peres,PPT} is based on the partial
transposition: if a state $\rho$ is separable then its partial
transposition $(\oper \ot \tau)\rho$ is positive. States which are
positive under partial transposition are called PPT states.
Clearly each separable state is necessarily PPT but the converse
is not true. It was shown by Horodecki et al. \cite{Horodeccy-PM}
that PPT condition is both necessary and sufficient for
separability for $2 \ot 2$ and $2 \ot 3$ systems.

Now, since all separable states belong to a set of PPT states, the
structure of this set is of primary importance in quantum
information theory.  Unfortunately, this structure is still
unknown, that is,  one may check whether a given state is PPT but
we do not know how to construct a general quantum state with PPT
property.  There are several well known examples of PPT states.
One class contains PPT states which are separable, e.g. Werner
\cite{Werner1} and isotropic states \cite{Horodecki}. Other
examples presents PPT states that are entangled. Actually there is
a systematic method of construction of PPT entangled states which
is based on a concept of unextendible product bases \cite{UPB}
(see also \cite{UPB-inne}). Other examples of PPT entangled states
were constructed in \cite{PPT,Horodecki-book,Shor,Ex2,Ex3,Ex4}.
PPT states play also a crucial role in mathematical theory of
positive maps and, as is well know, these maps are very important
in the study of quantum entanglement.  Recently, the mathematical
structure of quantum states with positive partial transposition
were studied in \cite{Ha,Kye}.

In the present Letter we propose a new class of bipartite $d \ot
d$ PPT states. Why this class is important: (i) it contains many
above mentioned examples of  PPT states. (ii) We claim that this
is the most general class of PPT states available at the moment.
Moreover, unlike other examples it fully uses complex
parametrization of density operators. (iii) Finally, it may be
used to study important properties of positive maps, e.g. to test
whether a given positive map is indecomposable and atomic. As is
well know indecomposable positive maps are crucial in constructing
entanglement witnesses.

The defining property of this class is very simple: it contains
bipartite states  invariant under the maximal commutative subgroup
of $U(d)$, i.e. $d$-dimensional torus $T^d=U(1) \times \ldots
\times U(1)$. This commutative subgroup is generated by $d$
mutually commuting operators
\begin{equation}\label{}
    \hat{t}_k = |k\>\<k| \ , \ \ \ \ k=1,\ldots,d \ ,
\end{equation}
where $|k\>$ denotes an orthonormal base in $\Cd$. Now, any vector
$\mathbf{x} \in \mathbb{R}^d$ gives rise to the following element
from $T^d$:
\begin{equation}\label{}
   U_\mathbf{x}  = e^{-i \mathbf{x}\cdot \widehat{\mathbf{t}}}\ ,
\end{equation}
where $\widehat{\mathbf{t}}= (\hat{t}_1,\ldots,\hat{t}_d)$.
Evidently, $U_\mathbf{x}U_\mathbf{y}=U_{\mathbf{x}+\mathbf{y}}$.

There are two classes of bipartite states invariant under
$U_\mathbf{x}$:

\begin{enumerate}

\item Werner-like state, or $U_\mathbf{x}\ot U_\mathbf{x}$--invariant states

\begin{equation}\label{}
    U_\mathbf{x} \ot U_\mathbf{x}\,\rho = \rho\, U_\mathbf{x}\ot U_\mathbf{x}\ ,
\end{equation}

\item isotropic-like state, or $U_\mathbf{x}\ot {U_\mathbf{x}^*}$--invariant states

\begin{equation}\label{}
    U_\mathbf{x} \ot {U_\mathbf{x}^*}\,\rho = \rho\, U_\mathbf{x}\ot {U_\mathbf{x}^*}\ ,
\end{equation}
for all $\mathbf{x} \in \mathbb{R}^d$. ${U_\mathbf{x}^*}$ denotes
complex conjugation of $U_\mathbf{x}$ in a fixed basis.

\end{enumerate}
Clearly, these two classes are related by a partial transposition,
i.e. a bipartite operator $\hat{O}$ is $U_\mathbf{x}\ot
{U_\mathbf{x}^*}$--invariant iff $(\oper \ot \tau)\hat{O}$ is
$U_\mathbf{x}\ot U_\mathbf{x}$--invariant.

The most general state which is $U_\mathbf{x}\ot
U_\mathbf{x}^*$--invariant  has the following form:
\begin{equation}\label{U---U}
    {\rho} = \sum_{i,j=1}^d a_{ij} \, |ii\>\<jj| + \sum_{i\neq j
    =1}^d c_{ij}\, |ij\>\<ij|\ .
\end{equation}
Now, since $\rho^\dag = \rho$  the matrix $\widehat{a}=||a_{ij}||$
has to be hermitian and $d^2-d$ coefficients $c_{ij}$ have to be
real. Moreover, $\rho$ is positive iff
\begin{equation}\label{CI}
    \widehat{a}=||a_{ij}|| \geq 0 \  \ \ \mbox{and}\  \ \ c_{ij} \geq 0 \ .
\end{equation}
Finally,  normalization $\mbox{Tr}\rho =1$ leads to
\begin{equation}\label{Tr=1}
    \mbox{Tr}\, \widehat{a} + \sum_{i\neq j
    } c_{ij} =1\ .
\end{equation}
Consider now the partial transposition of $\rho$:
\begin{equation}\label{}
   (\oper \ot \tau) \rho = \sum_{i,j=1}^d {a_{ij}} \, |ij\>\<ji| + \sum_{i\neq j
    =1}^d c_{ij}\, |ij\>\<ij|\ .
\end{equation}
Note, that the above formula  may be rewritten as follows
\begin{equation}\label{}
    (\oper \ot \tau)\rho = \sum_{i=1}^d a_{ii}\, |ii\>\<ii| + \sum_{i<j} \widehat{X}_{ij}\ ,
\end{equation}
where the  operator $\widehat{X}_{ij}$ is given by
\begin{equation*}\label{}
    \widehat{X}_{ij} = a_{ij} |ij\>\<ji| +
    {a_{ij}^*}|ji\>\<ij| + c_{ij}|ij\>\<ij| +
    c_{ji}|ji\>\<ji| \ .
\end{equation*}
Since two operators $\sum_{i} a_{ii}\, |ii\>\<ii|$ and $\sum_{i<j}
\widehat{X}_{ij}$ live in a mutually  orthogonal subspaces of $\Cd
\ot \Cd$, the positivity of $(\oper \ot \tau)\rho$ implies
separately $\sum_{i} a_{ii}\, |ii\>\<ii|\geq 0$, which is
equivalent to $a_{ii}\geq 0$, and $\sum_{i<j} \widehat{X}_{ij}\geq
0$. Now, for any pair $i< j$ an operator $\widehat{X}_{ij}$ acts
on a 2-dimensional subspace of $\Cd \ot \Cd$ spanned by $|ij\>$
and $|ji\>$:
\begin{eqnarray}\label{aacc}
\widehat{X}_{ij}|ij\> &=&   c_{ij} |ij\>  + a_{ij}|ji\> \ , \nonumber\\
\widehat{X}_{ij}|ji\> &=& {a_{ij}^*}|ij\> + c_{ji} |ji\> \ ,
\end{eqnarray}
and hence $\widehat{X}_{ij}|ij\> \geq 0$ iff
\begin{equation}\label{}
\left[ \begin{array}{cc} c_{ij} & a_{ij} \\ {a_{ij}^*} & c_{ij}
\end{array} \right] \, \geq\, 0\ ,
\end{equation}
which is equivalent to the following condition
\begin{equation}\label{CII}
    c_{ij}c_{ji} - |a_{ij}|^2 \geq 0\ .
\end{equation}
There is an evident example  of isotropic-like PPT states. Let
$\overrightarrow{\lambda}=(\lambda_1,\ldots,\lambda_d)$ be a
normalized complex vector. Consider a state from a class
(\ref{U---U}) with $a_{ij} = \lambda_i {\lambda}^*_j$ and $c_{ij}
= |\lambda_i {\lambda}^*_j|$. Evidently $\widehat{a}\geq 0$,
$c_{ij}\geq 0$, and
 $   c_{ij}c_{ji} - |a_{ij}|^2 = 0\ .$
Hence each complex vector $\overrightarrow{\lambda}$ gives rise to
a PPT state.

Similarly one may analyze a general Werner-like $U_\mathbf{x} \ot
{U_\mathbf{x}^*}$--invariant state:
\begin{equation}\label{UU}
    \widetilde{\rho} = \sum_{i,j=1}^d b_{ij} \, |ij\>\<ji| + \sum_{i\neq j
    =1}^d c_{ij}\, |ij\>\<ij|\ .
\end{equation}
Now, positivity of $\widetilde{\rho}$ is equivalent to $c_{ij}\geq
0$ and
\begin{equation}\label{CIII}
    c_{ij}c_{ji} - |b_{ij}|^2 = 0\ .
\end{equation}
On the other hand partial transposition
\begin{equation}\label{}
   (\oper \ot \tau) \widetilde{\rho} = \sum_{i,j=1}^d {b_{ij}} \, |ii\>\<jj| + \sum_{i\neq j
    =1}^d c_{ij}\, |ij\>\<ij|\ ,
\end{equation}
is $U_\mathbf{x}\ot U_\mathbf{x}^*$--invariant and, therefore,
$\widetilde{\rho}$ is PPT iff
\begin{equation}\label{CIV}
\widehat{b}=||b_{ij}|| \geq 0 \  \ \ \mbox{and}\  \ \ c_{ij} \geq
0 \ .
\end{equation}

\vspace{.2cm}

\noindent {\it Examples} --- Now we show that many well known
examples of PPT states belong to our class.

{\it 1. Werner state} \cite{Werner1}:
\begin{equation}\label{W}
    \mathcal{W}_p = (1-p) Q^+ + pQ^-\ ,
\end{equation}
where
\begin{equation*}\label{}
    Q^\pm = \frac{1}{d(d\pm 1)} \Big( I \ot I \pm \sum_{i,j=1}^d |ij\>\<ji|
    \Big) \ .
\end{equation*}
Clearly, ${\cal W}_p$ belongs to a class (\ref{UU}) with
\begin{equation*}\label{}
    b_{ij} = \left\{\begin{array}{ll} x_-\ \ & , \ i\neq j \\
 x_-    + x_+ \ \ &, \ i=j \end{array} \right. \ ,
\end{equation*}
and $c_{ij} = x_+$, where
\begin{equation*}\label{}
    x_\pm = \frac{1-p}{d^2 + d} \pm \frac{p}{d^2 - d} \ .
\end{equation*}
Condition (\ref{CIII}) implies  $p\in [0,1]$. To check condition
(\ref{CIV}) for PPT let us observe that the spectrum of
$\widehat{b}$ consists of only two points: $\lambda_1=x_+$ with
muliplicity $d-1$ and $\lambda_2=dx_- + x_+$. Therefore
$\widehat{b}$ is positive iff $\lambda_2\geq 0$ which is
equivalent to $p\leq 1/2$ and hence it reproduces well known
result for PPT property of Werner states \cite{Werner1}.

{\it 2. Isotropic state} \cite{Horodecki}:
\begin{equation}\label{}
    \mathcal{I} = \frac{1-\lambda}{d^2}\,I \ot I + \frac{\lambda}{d}\,
    \sum_{i,j=1}^d|ii\>\<jj|\ .
\end{equation}
belongs to a class (\ref{U---U}) with $ c_{ij} = (1-\lambda)/d^2$
and
\begin{equation*}\label{}
   a_{ij} = \left\{\begin{array}{ll}
 \lambda/d\ \ & , \ i\neq j \\ \lambda/d + (1-\lambda)/d^2 \ \ &, \ i=j \end{array} \right. \
 .
\end{equation*}
Positivity of $\widehat{a}$ together with $c_{ij}\geq 0$ imply
 $   -1/(d^2-1)\leq  \lambda \leq 1$.
PPT condition (\ref{CII}) leads to $\lambda\leq 1/d+1$ which
reproduces well known result for PPT property of isotropic states
\cite{Horodecki}.

 {\it 3.} Authors of
 \cite{Shor} considered the following two-parameter family
\begin{eqnarray}\label{}
    \rho_{bc} &=& a\,\sum_{i=1}^d|ii\>\<ii| + b
    \sum_{i<j=1}^d|\psi^-_{ij}\>\<\psi^-_{ij}| \nonumber \\ &+& c
    \sum_{i<j=1}^d|\psi^+_{ij}\>\<\psi^+_{ij}| \ ,
\end{eqnarray}
where $\, |\psi^\pm_{ij}\> =  (|ij\> \pm |ji\> )/\sqrt{2}\,$.
Note, that the unit trace condition (\ref{Tr=1}) enables one to
compute $a$ in terms of $b$ and $c$: $a =  1/d - (b+c)/(2d-2)$.
Clearly, $\rho_{bc}$ belongs to a  Werner class (\ref{UU}) with
\begin{equation*}\label{}
   b_{ij} = \left\{\begin{array}{ll}
 (c-b)/2\ \ & , \ i\neq j \\ a \ \ &, \ i=j \end{array} \right. \
 ,
\end{equation*}
and $c_{ij} = (c+b)/2$. Now, $c_{ij}\geq 0$ implies $c+b\geq 0$
whereas $\widehat{b}\geq 0$ gives:
\begin{eqnarray*}\label{}
    1 - d(d-1)b & \geq & 0 \ , \\
    2 - d(d-2)\, b - d^2\, c & \geq& 0 \ ,
\end{eqnarray*}
which reproduce results of \cite{Shor}.

{\it 4. } Horodecki {\em et al.} \cite{Horodecki-book} considered
the following $3 \ot 3$ state
\begin{equation}\label{HOR}
    \sigma_\alpha = \frac 27 P^+ + \frac \alpha 7\, \sigma_+ +
    \frac{5-\alpha}{7}\,\sigma_-\ , \ \ \ 2\leq\alpha\leq 5\ ,
\end{equation}
where $P^+$ denotes a projector onto the canonical maximally
entangled state, and
\begin{eqnarray*}\label{}
    \sigma_+ &=& \frac 13\, ( |01\>\<01| + |12\>\<12| + |20\>\<20|)
    \ , \\
\sigma_- &=& \frac 13\, ( |10\>\<10| + |21\>\<21| + |02\>\<02|)
    \ .
\end{eqnarray*}
 Clearly, (\ref{HOR})
belongs to isotropic-like class (\ref{U---U}) with $a_{ij} =
2/21$,  $c_{01}=c_{12}=c_{20} = \alpha/21$ and $
c_{10}=c_{21}=c_{02} = (5-\alpha)/21$. One easily finds that PPT
condition (\ref{CII}) reproduces well known fact that (\ref{HOR})
is PPT for $\alpha\leq 4$. Recently, a one-parameter family
(\ref{HOR}) was generalized for $d \ot d$ systems as follows
\cite{Ex4}:
\begin{equation}\label{}
    \rho = \frac{a_1}{d} P^+ + \sum_{i=1}^d\sum_{j=2}^d\,
    \frac{a_j}{d} |i,i+j-1\>\<i,i+j-1|\ ,
\end{equation}
where the positive numbers $a_i$ satisfy $\sum_i a_i =1$. Clearly,
it belongs to an isotropic-like family (\ref{U---U}). If
$a_{i+1}a_{d-i+1}\geq a_1^2$ then the state is PPT.

{\it 5.} Bound entangled states considered in \cite{Ex2} belong to
our class (\ref{U---U}) with $a_{ij}=1$.

{\it 6. St{\o}rmer state}. St{\o}rmer \cite{Stormer} analyzed an
un-normalized $3 \ot 3$ positive  PPT matrix with $a_{ij}=1$ and
\begin{equation*}\label{}
 c_{ij}= 2\mu \ , \ \ i<j \ ;\ \  c_{ij} =
    1/2\mu \ ,\ \ \ i>j  \ .
\end{equation*}
This example may be immediately generalized for $d \ot d$ case as
follows: $a_{ij} = \alpha$,  and
\begin{equation*}\label{}
c_{ij} >0 \ , \ \ i<j \ ;\ \  c_{ij} = \alpha^2\, c^{-1}_{ji} \ ,\
\ \ i>j  \ ,
\end{equation*}
where $\alpha>0$ is a normalization constant. One has $\,
c_{ij}c_{ji} = |a_{ij}|^2$, and (\ref{CII}) implies that the
corresponding state is PPT.

{\it 7. } Ha \cite{Ha} performed very sophisticated construction
of a one-parameter family of $d^2 \times d^2$ (un-normalized)
positive matrices and showed that this family remains positive
after performing partial transposition. It turns that Ha's family
is a special example of an isotropic-like class (\ref{U---U}) with
$a_{ij} =1$ and
\begin{equation*}\label{}
    c_{i\oplus 1,i} = \lambda\ , \ \ \ c_{i,i\oplus 1} = \lambda'\ ,
\end{equation*}
and the remaining $c_{ij}=1$. In the above formula `$\oplus$'
denotes addition modulo $d$, and
\begin{equation*}\label{}
    \lambda = \frac{\gamma^2 + d-1}{d}\ , \ \ \ \lambda' = \frac{\gamma^{-2} + d-1}{d}\
    ,
\end{equation*}
with  $\gamma >0$. Now, conditions for positivity (\ref{CI}) are
trivially satisfied. Moreover, due to $\lambda\lambda' \geq 1$,
the PPT condition (\ref{CII}) is also satisfied which shows that
Ha's family is PPT.

Another example constructed in \cite{Ha} is a family of
(un-normalized) $3\ot 3$ positive PPT matrices but the
construction may be generalized to an arbitrary $d$ as follows:
let
\begin{eqnarray*}\label{}
u_i &=& |ii\> \ ,\ \ \ \     z_i = \frac 1s\, |i+1,i\> + s\,
|i,i+1\> \ ,
\end{eqnarray*}
with $s>0$ and $i=1,\ldots,d$. Define the following family of
positive $d^2\times d^2$ matrices:
\begin{equation}\label{Bs}
    B_s = \sum_{i=1}^d \Big( |u_i\>\<u_i| +  |z_i\>\<z_i|\Big)\ .
\end{equation}
Observe, that (\ref{Bs}) belongs to a Werner-like class (\ref{UU})
with
\[ b_{ii} = b_{i,i\oplus 1} = b_{i\oplus1 ,i} =1\ , \]
\[ c_{i,i\oplus 1} = s^2\ , \ \ \ \ c_{i\oplus 1,i} = s^{-2}\ , \]
and the remaining $b_{ij}$ and $c_{ij}$ vanish. Note, that PPT
condition (\ref{CIV}) is trivially satisfied.

{\it 8. PPT states which do not belong to our class}. It turns out
that apart from bound entangled states constructed {\em via}
unextendible product bases \cite{UPB} almost all other examples of
PPT states do belong to our class. We are aware of only few
exceptions:  one is the family  of $O\ot O$--invariant states
\cite{Werner2,II} and the second one is the celebrated family of
$3 \ot 3$ states which are nonseparable but PPT constructed by
Horodecki \cite{PPT}. Note that the $d$-dimensional torus $U(1)
\times \ldots \times U(1)$ does not allow for an orthogonal
subgroup and hence states with orthogonal symmetry has to be
considered separately. Now, the Horodecki state $\rho_a$ may be
rewritten as
  $  \rho_a = \rho'_a + \rho''_a$,
where
\begin{equation*}\label{}
    \rho'_a = \alpha a\,\Big( \sum_{i,j=1}^3 |ii\>\<jj| +
    \sum_{i\neq j =1}^3\, |ij\>\<ij| \Big)\ ,
\end{equation*}
and
\begin{equation*}\label{}
    \rho''_a = \frac{\alpha}{2}\, \sqrt{1-a^2}\, (  |31\>\<33| +
    |33\>\<31|)\ ,
\end{equation*}
with $\alpha = 1/(8a+1)$ being a normalization constant. Note,
that $\rho'_a$ is an isotropic-like matrix and does belong to
(\ref{U---U}). However, $\rho''_a$ is not invariant under the
maximal commutative subgroup of $U(3)$. Note, that $\rho''_a$ is
invariant only under 1-parameter subgroup generated by
$\hat{t}_2=|2\>\<2|$. It shows that Horodecki state $\rho_a$ is
also symmetric but with respect to smaller symmetry group.

\vspace{.2cm}

\noindent {\em Separability.} --- A state from a class
(\ref{U---U}) is separable iff there exists a separable state
$\sigma$ such that $\rho=\mathbf{P}\sigma$, where $\mathbf{P}$
denotes a projector operator projecting an arbitrary state onto
the class (\ref{U---U}), i.e. $\mathbf{P}\sigma$ belongs to
(\ref{U---U}) with `pseudo-fidelities'  $a_{ij} = \mbox{Tr}(\sigma
|ii\>\<jj|)$ and $c_{ij} = \mbox{Tr}(\sigma |ij\>\<ij|)$. Taking
$\sigma = |\alpha \ot \beta\>\<\alpha\ot \beta|$ one finds the
following sufficient condition for separability:
\begin{eqnarray}  \label{SEP}
 a_{ij} = {\alpha^*_i}\alpha_j{\beta^*_i}\beta_j\ , \ \ \ \ \
c_{ij} = |\alpha_i|^2|\beta_j|^2\ ,
\end{eqnarray}
where $\alpha_i = \<i|\alpha\>$ and $\beta_i=\<i|\beta\>$.

\vspace{.2cm}

\noindent {\em Positive maps.} --- PPT states are also important
in the study of positive maps \cite{Stormer,Ha,MAPS} (see also
\cite{Horodecki-book} for a useful review). It has been shown
\cite{Horodeccy-PM} that there exists a strong connection between
the classification of the entanglement of quantum states and the
structure of positive linear maps. Let $M_d$ denote a set of $d
\times d$ complex matrices and $V_k$ be a cone of positive
matrices  $A \in (M_d \ot M_d)^+$  such that Schmidt number
$\mbox{SN}(A) \leq k$ \cite{SN}. Now, one says that $A$ belongs to
a cone $V^l$ iff $A$ is PPT and $\mbox{SN}[(\oper \ot \tau)A]\leq
l$. It is clear that $V_1=V^1$ defines a cone of separable
elements.

Recall that a positive map $\Phi : M_d \longrightarrow M_d$ is
$k$-positive iff $(\oper \ot \Phi)$ is positive when restricted to
$V_k$. Similarly, $\Phi$ is $k$-copositive iff $(\oper \ot
\Phi\circ \tau)$ is positive on $V^k$. The most basic class of
positive maps consists of so called atomic ones \cite{atomic} --
$\Phi$ is {\it atomic} iff it cannot be decomposed into the sum of
2-positive and 2-copositive maps. Atomic maps  posses the
`weakest' positivity property and hence may be used to detect the
bipartite states with the `weakest' entanglement, i.e.  states
from $V_2 \cap V^2$.  Conversely, PPT states may be used to check
the atomic property of positive maps. Suppose that we are given an
indecomposable positive map $\Phi$. If for some $A \in V_2 \cap
V^2$ one finds that $(\oper \ot \Phi)A$ is not positive then
$\Phi$ is necessarily atomic. Now, it would be interesting to know
when PPT states from our class belong to $V_2 \cap V^2$. Consider
e.g. a state $\rho$ from an isotropic-like class (\ref{U---U}).
Note, that if $\rho$ is PPT then, due to (\ref{aacc}), $\rho$
necessarily belongs to $V^2$. Hence, it is enough to check when
$\rho \in V_2$. It is clear that $\rho \in V_2$ iff there is
$\sigma' \in V_2$ such that $\rho = \mathbf{P}\sigma'$. Taking
$\sigma' = \frac 12\,|\alpha \ot \beta + \psi\ot \phi\>\<\alpha\ot
\beta + \psi \ot \phi|$ one finds the following sufficient
condition for $\rho$ to be an element form a cone $V_2$:
\begin{eqnarray*}\label{V2}
 a_{ij} &=& \frac 12 \Big( {\alpha^*_i\beta^*_i}[ \alpha_j\beta_j
 + \psi_j\phi_j ] + {\psi^*_i\phi^*_i}[ \alpha_j\beta_j
 + \psi_j\phi_j ]\Big) \ , \nonumber \\
 c_{ij} &=& \frac 12 \Big( {\alpha_i\beta_j}[ \alpha^*_i\beta^*_j
 + \psi^*_i\phi^*_j ] + {\psi_i\phi_j}[ \alpha^*_i\beta^*_j
 + \psi^*_i\phi^*_j ]\Big) \ ,
\end{eqnarray*}
where $\alpha_i = \<i|\alpha\>$ and similarly for $\beta_i,\psi_i$
and $\phi_i$.  Interestingly,  any Werner-like state from
(\ref{UU}) belongs to $V_2$. Hence it suffices  to check wether it
belongs to $V^2$. One may easily derive sufficient conditions for
$b_{ij}$ and $c_{ij}$ in analogy to the above conditions for
$a_{ij}$ and $c_{ij}$.

{\em Acknowledgments} ---  This work was partially supported by
the Polish State Committee for Scientific Research Grant {\em
Informatyka i in\.zynieria kwantowa} No PBZ-Min-008/P03/03.

\end{document}